\documentclass[12pt]{article}

\usepackage{setspace,graphicx,epstopdf,amsmath,amsfonts,amssymb,amsthm,versionPO}
\usepackage{booktabs}
\usepackage[ruled, lined, linesnumbered, commentsnumbered, longend]{algorithm2e}
\usepackage{marginnote,datetime,enumitem,subfigure,rotating,fancyvrb}
\usepackage{hyperref,float}
\usepackage{mathtools}
\usepackage{dsfont}
\usepackage{threeparttable}
\usepackage{tabularx}
\usepackage{siunitx}
\usepackage{natbib}
\usepackage{array,xltabular,threeparttablex}
\usepackage{tabu}
\usepackage{caption}
\usepackage{csquotes}
\usdate
\DeclareMathOperator*{\argmax}{arg\,max}
\DeclareMathOperator*{\argmin}{arg\,min}

\DeclarePairedDelimiter\floor{\lfloor}{\rfloor}

\newcolumntype{d}{D{.}{.}{4}}
\usdate

\excludeversion{notes}        
\includeversion{links}          

\iflinks{}{\hypersetup{draft=true}}

\ifnotes{%
\usepackage[margin=1in,paperwidth=10in,right=2.5in]{geometry}%
\usepackage[textwidth=1.4in,shadow,colorinlistoftodos]{todonotes}%
}{%
\usepackage[margin=1in]{geometry}%
\usepackage[disable]{todonotes}%
}

\makeatletter\let\chapter\@undefined\makeatother 

\makeatletter
\newcommand{\@giventhatstar}[2]{(#1\;\middle|\;#2)}
\newcommand{\@giventhatnostar}[3][]{#1(#2\;#1|\;#3#1)}
\newcommand{\giventhat}{\@ifstar\@giventhatstar\@giventhatnostar}
\makeatother

\usepackage{indentfirst}
\usepackage{jfe}

\begin{document}

\setlist{noitemsep}

\title{Advanced Statistical Learning on Short Term Load Process Forecasting\footnotetext{Supported by Deutsche Forschungsgemeinschaft as a transfer project collaborated with \"Okotec GmbH via Humboldt-Universit\"at zu Berlin.}}


\author{Junjie Hu, Brenda L\'opez Cabrera, Awdesch Melzer\\
  Humboldt-Universit\"at zu Berlin}

\date{03.10.2021}

\renewcommand{\thefootnote}{\fnsymbol{footnote}}

\singlespacing

\maketitle

\vspace{-.2in}
\begin{abstract}
\noindent 
Short Term Load Forecast (STLF) is necessary for effective scheduling, operation optimization trading, and decision-making for electricity consumers. Modern and efficient machine learning methods are recalled nowadays to manage complicated structural big datasets, which are characterized by having a nonlinear temporal dependence structure. We propose different statistical nonlinear models to manage these challenges of hard type datasets and forecast 15-min frequency electricity load up to 2-days ahead. We show that the Long-short Term Memory (LSTM) and the Gated Recurrent Unit (GRU) models applied to the production line of a chemical production facility outperform several other predictive models in terms of out-of-sample forecasting accuracy by the Diebold-Mariano (DM) test with several metrics. The predictive information is fundamental for the risk and production management of electricity consumers.

\end{abstract}

\medskip



\thispagestyle{empty}

\clearpage

\onehalfspacing
\setcounter{footnote}{0}
\renewcommand{\thefootnote}{\arabic{footnote}}
\setcounter{page}{1}

\section{Introduction}

Accurate short term load forecast (STLF) is the key component in energy management, and more and more energy-intensive companies have been demanding new methodologies to reduce their daily costs. This paper is a collaboration between academic research and a real-world case from the industry. So far, much of the literature studies the short term load forecast dealing only with the cases of the aggregated level of electricity load. To our knowledge, there is no study on forecasting the machine-scale electricity consumption, and we propose linear/nonlinear models on a load process generated from a chemical facility. We model this complicated load process by adjusting the restriction on parameters using three different methods, including the classic parametric AR-type model, a semi-parametric model, and state-of-the-art machine learning models.

Short term load forecast on small-scale energy consumers, such as a machine or a household, is not commonly seen in previous literature but getting crucial nowadays. Unlike the load processes in aggregated levels which are often characterized by certain seasonality, the electricity load from a small-scale consumer always evolves with abrupt jumps and incorporates noise. Therefore, the pattern of such a load process is almost unrecognizable. Nevertheless, forecast on the small-scale consumer is demanded in the industry, notwithstanding the difficulties of such modeling. Companies can better optimize their production schedules with the knowledge about the energy consumption of each one of their machines, and retail consumers can also implement the forecasting models as the energy trading market is now expanding to households in many countries thanks to the rapid development of demand-side management (\cite{Weber_2017}). This study fits the gap and discusses the predictability of such load processes.

The load process modeled in this paper is characterized by having a complicated structure. More specifically, the dataset we used is filled with jumps or structure breaks and various types of noises. The jumps are generally modeled as a c\`adl\`ag process, i.e., the jumps are not predictable without extra information. Such a structure is a typical situation of modeling small-scale energy consumers where exogenous shocks become dominant. Papers discuss the forecasts of the household electricity consumption as the development of smart grids in recent years extensively, for example, \cite{Amara2017} propose an adaptive model to forecast the conditional density of household electricity demand.

Load forecast is a major research topic in energy fields. In terms of time-scale, the literature can be divided into the long term forecast, e.g., monthly load forecast (\cite{chang2011monthly}), and the short term forecast, which is our focus. In terms of individual-scale, the existing literature mainly studies load process aggregated in a large-scale, such as the state-level (\cite{bianco2009electricity, darbellay2000forecasting,do2016electricity,ougcu2012forecasting,Castelli_2015,Pardo_2002}). All the load processes in those articles tend to have clear patterns of seasonalities. We obtained a unique dataset consisted of an intraday electric load process with the corresponding production schedules from a chemical facility. 
On each day, we forecast a 15-minute frequency load process in the following two days.

In our case, one question is how much the recent advanced models can outperform the classic time-series models. One can find literature applying neural network models on STLF,
such as \cite{an2013using, bianchi2017overview, guan2012very} concluding that modeling nonlinearity does improve the forecasting accuracy. However, the question remains in our case that does not appear with clear cyclic patterns. Hence a classic mode decomposition or noise filtering method is unlikely useful. Many other different approaches have been applied to the STLF. A strand of literature is motivated to describe the seasonality with the single/multiple equation(s) models or functional data analysis (\cite{taylor2003short,cottet2003bayesian,bianchi2017overview, clements2016forecasting, Schulz:2014}). Most of those models are carefully designed and calibrated. However, they do not fit in our case as they are restricted to certain fixed seasonal patterns. Another angle to improve the forecasting accuracy is by accounting for the nonlinear dependence of the load processes using semi-parametric or non-parametric models. 
We show that the Long-short Term Memory (LSTM) and the Gated Recurrent Unit (GRU) architectures outperform a semi-parametric functional data method named FActorisable Sparse Tail Event Curves (FASTEC) (\cite{Chao:2015b}) and the ARX model. The forecast does show a consistent result with the literature. However, LSTM or GRU models should be carefully used since it losses forecasting power on tails or jumps, which we believe is the key conclusion that should be emphasized for many practitioners, i.e., the machine learning is \emph{NOT} omnipotent. 

The remainder of this paper is organized as follows. Section \ref{sec:load process} gives an overview of the Data of a chemical production facility and conducts the empirical analysis of the high frequency data. In section \ref{forecasting}, we discussed different forecast models, statistical and machine learning models, and evaluate their performance. Section \ref{conclusion} concludes and make suggestions for further research. All computations for this study were carried out in Python.

\section{Data and stylised facts} \label{sec:load process}

The dataset is provided by the consulting company \"Okotec Energiemanagement GmbH and contains the total electricity consumption and 72 production schedules of a product line from one chemical production facility of an anonymous chemical company in Germany. The time series ranges from 1st January 2017 to 31st December 2017 and contains 96 observations daily, i.e., a 15-minute resolution, and the 72 production schedules can be divided into a set of 3 major groups of production schedules. Moreover, the production schedules are lagging for 2 hours, meaning that the information about the production schedules is saved 2 hours after the actual production process has started. Hence, the electricity consumption process is matched at the 8th lag with all the production schedules under the assumption that production schedules are known contemporaneously.

\begin{figure}[!htb]
  \centering
  \begin{minipage}{1\linewidth}
    \includegraphics[width=\textwidth]{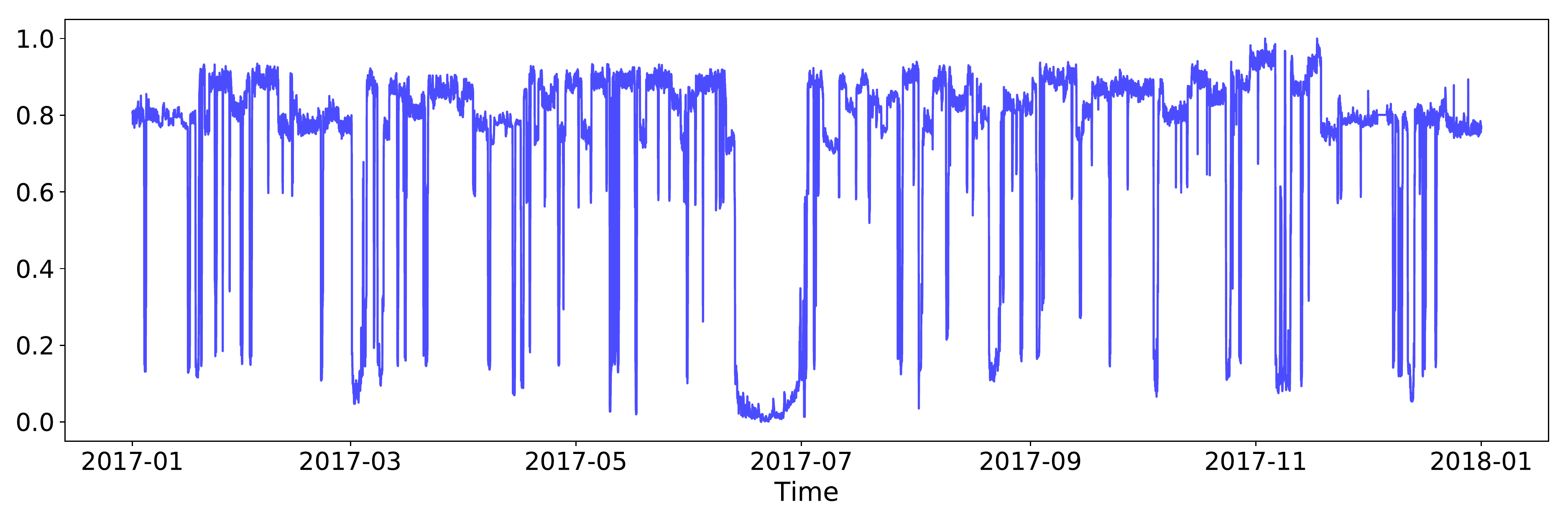}
  \end{minipage}
  \caption{
    \textbf{Electricity Load Process}
  \newline 
  \small
  The figure shows the 15-minute frequency electricity consumption process from 1.Jan.2017 to 31.Dec.2017. The load process is scaled down into [0,1].
  }\label{fig:load_ts}
\end{figure}

The load process is scaled into [0,1] range, shown in Figure \ref{fig:load_ts} in which we can directly see that the frequent jumps on the load process are caused by switching between different products or fully-loaded/minimum-loaded. Hence, then we can see that machinery is working at different states at different periods.

Another issue of the data the dimensionality. Having detailed information about the production schedule of all 72 product lines is informative for the producer but leads to inefficiency and inconsistent estimators in a time series regression model due to the large coefficient space. There exist several methods that deal with the problem. Regularized regressions techniques, such as elastic net, scad, ridge regression, and LASSO, avoid this curse of dimensionality and shrink the parameter space by using data aggregation. However, they can be slow and not performing well compared to ARMA or linear regression models. Moreover, given rigid server-time constraints by the consulting company, our production schedules had to be burned down to an aggregate of three groups. Large-scale energy consumers have to give their bids and orders at the energy market every day until 2 pm. By 10 am the current data is fed to the consulting company.

(Weak) Stationarity is one of the prerequisites for econometric models calibration, analysis, and inference. We found the load process interesting as two of the most commonly used stationarity tests, the Augmented Dickey-Fuller (ADF, \cite{dickey1979distribution,dickey1981likelihood}) test and Kwiatkowski–Phillips–Schmidt–Shin (KPSS, \cite{kwiatkowski1992testing}) test, tell differently.

\begin{table}[htb]
\centering
  \small
  \setlength{\tabcolsep}{0pt}
  \begin{threeparttable}
    \caption{Stationarity Test}\label{tab:adf_kpss}
    \begin{tabular*}{0.7\linewidth}{@{\extracolsep{\fill}}>{\itshape}l *{2}{S[table-format=2.2,table-number-alignment = center]} c}
        \toprule\toprule
         & \multicolumn{1}{c}{test} & \multicolumn{1}{c}{$p$-value} & \multicolumn{1}{c}{lags} \\
        \cline{2-4}
    ADF  & -10.79 & 0.00 & 51 \\
    KPSS &  0.83 &  0.01 & 52 \\

          \bottomrule\bottomrule
        
    \end{tabular*}
  \begin{tablenotes}[flushleft]
    \setlength\labelsep{0pt}
  \linespread{1}\small
  \item 
    The table reports the results of two stationarity tests, ADF and KPSS, on the electricity load process. Both tests suggest rejecting their null hypotheses, implying a conflicted conclusion from the stationarity test.
  \end{tablenotes}
  \end{threeparttable}
\end{table}

The null hypothesis of ADF is unit root exists in the process against the alternative that such process is (trend) stationary. On the other hand, the null hypothesis of KPSS is that the observed process is (trend) stationary against the unit root alternative. The results from Table \ref{tab:adf_kpss} reject both null hypothesis under the 1\% significant level. We may conclude that the load process is neither stationary nor unit root process. Lack of stationarity could cause most of the estimation for statistical inference to be biased or inconsistent. The reason for such contradictory might be a significant amount of structural breaks in the process, in which case the assumptions for tests do not hold. On the other hand, the process is by default stationary, as the maximum electricity consumption level of that single machinery cannot increase over time.

In this article, we do not model the structural break explicitly. Modeling structural breaks depend heavily on detection, which is most likely to be ex-post methods. Literature shows that structural break doesn't affect forecast result significantly per se (\cite{stock1996evidence}). Also, the benefits of modeling structural break could be canceled out by the estimation errors of break time points and parameters (\cite{elliott2007confidence,elliott2014pre}). The commonly used forecast metric, Mean Squared Error (MSE), is always facing a bias-variance trade-off which means ignoring structural break could bring even more accurate forecast (\cite{pesaran2005small}).

\begin{figure}[!htb]
  \centering
  \begin{minipage}{1\linewidth}
    \includegraphics[width=\textwidth]{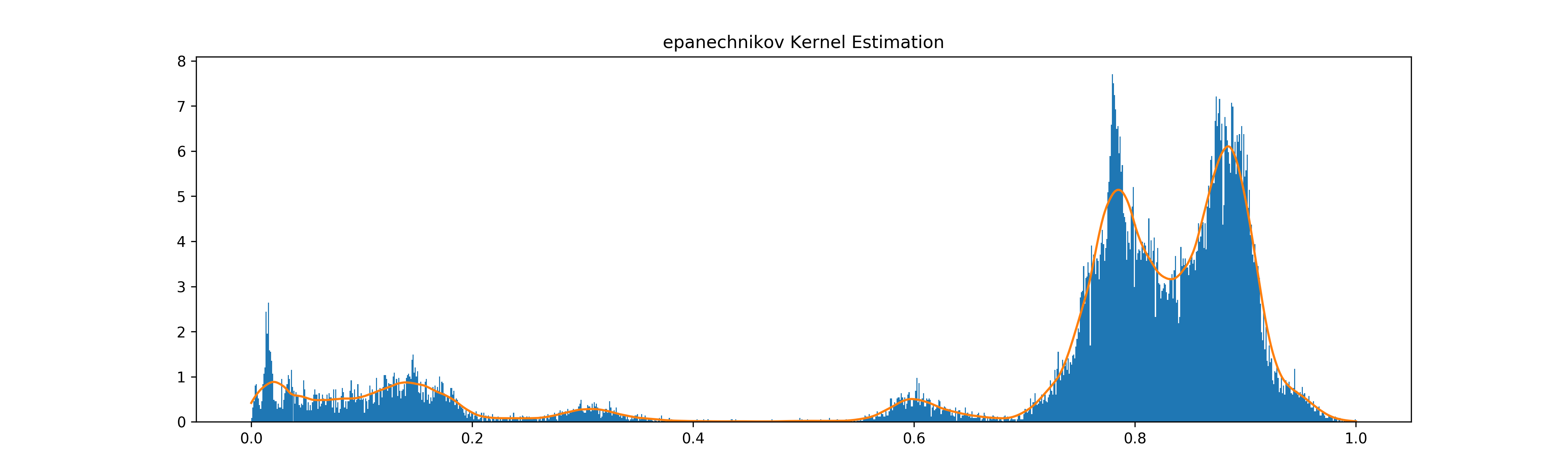}
  \end{minipage}
  \caption{
    \textbf{KDE of the Electricity Load Process}
  \newline 
  \small
Kernel density estimation of full sample electricity consumption values. Epanechinikov kernel with bandwidth 1.6\%.
  }\label{fig:load_kde}
\end{figure}

From Figure \ref{fig:load_kde}, we can see that the load is mixture-Gaussian distributed, implying that the load process consists of a certain number of states and stationary process within each state. The load process is characterized by a slow decay of ACF (Long-memory property), which enables us to forecast with lagged load values, and therefore the classic AR-type models can be applied.

\begin{figure}[!htb]
  \centering
  \begin{minipage}{1\linewidth}
    \includegraphics[width=\textwidth]{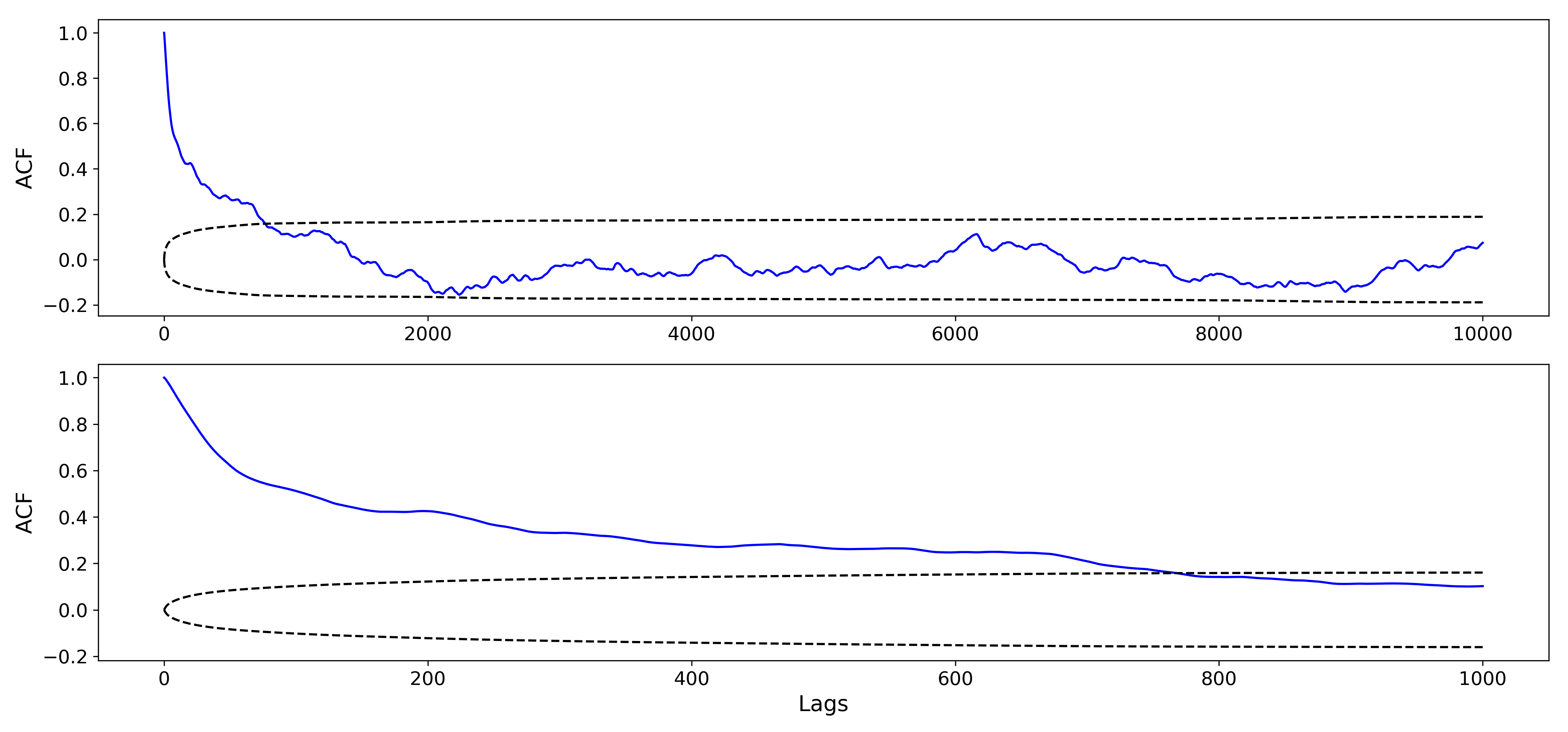}
  \end{minipage}
  \caption{
    \textbf{ACF of the Electricity Load Process}
  \newline 
  \small
The figure shows the Autocorrelation function of the load process. The upper panel shows 10000 maximum lags; the bottom panel has 1000 maximum lags. The solid line is the autocoefficient on different lags, and the dashed line forms the confidence interval at a 5\% significant level.
  }\label{fig:load_acf}
\end{figure}

\begin{figure}[!htb]
  \centering
  \begin{minipage}{1\linewidth}
    \includegraphics[width=\textwidth]{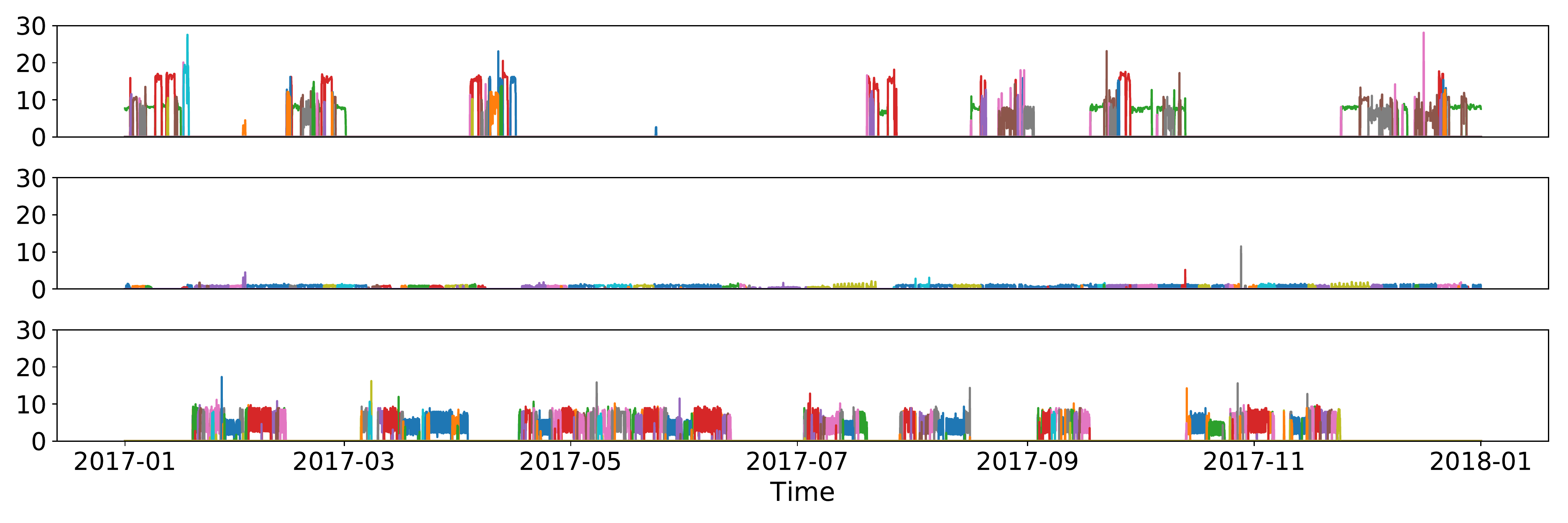}
  \end{minipage}
  \caption{
    \textbf{The 3-Group of Production Schedules}
  \newline 
  \small
The figures show the production schedules grouped into three groups (from top to bottom) based on the products being manufactured. The production schedules are recorded 2-hour late after the actual production. We shift the production schedules 2-hour earlier, assuming that the schedules are known as soon as they start.
  }\label{fig:prodction}
\end{figure}

Apart from the load process, this dataset also includes 72 sub-product schedules. There are three major products manufactured in this production line, and each schedule corresponds to one of those three products. Therefore, we divide those schedules into three groups as shown in Figure \ref{fig:prodction}. For further investigation, we aggregate those sub-product schedules into three groups as exogenous variables that can be considered as a relative volume for three independent products.

\section{Forecasting Models}\label{forecasting}

In this section, we first construct and justify the advantages of each forecasting model, followed by analyzing the forecast results with different metrics. Three approaches are taken in this article, including linear parametric model, semi-parametric model, and non-parametric model. The non-parametric approach, neural network models LSTM and GRU, are of our focus to capture the nonlinear temporal dependency of the process. In addition, a functional data model "FASTEC" combined with the regime-switching method detecting different states at each time $t$ is also constructed. Most of the load forecasting purposes are related to trading on the day-ahead energy market, which requires forecasts at least 36-hours ahead. Hence, the forecasting horizon of both models is 192-steps (2-day, 15-minute per step) ahead. In each day $t$, $t=1,2,\ldots,T$, The forecasting task can be generally written as,

\begin{equation}
    Y_{t+2} = g\left({\bf X}_{t}; \Theta\right),
\end{equation}

where the forecasting load with 192 steps, $Y_{t+2} \stackrel{\operatorname{def}}{=} \{y_{s,t+1}, y_{s,t+2}\}_{s=1}^{S}$, $y_{s,t+h}\in \mathbb{R}$, $h=1,2$;  and the predictor variables ${\bf X_{t}}\stackrel{\operatorname{def}}{=} \{X_{s,t}\}_{s=1}^{S}$ with feature vector $X_{s,t}\in \mathbb{R}^{k}$, and $k=4$. $S$ is fixed as 96 for the 15-minute sampling frequency each day.

For different forecasting models, function $g$ is chosen within a different class of functions, and the parameter $\Theta$ is calibrated accordingly. We demonstrate how to prepare the forecasting samples with the following algorithm 1 representation:

\begin{algorithm}
    \SetKwInOut{KwIn}{Input}
    \SetKwInOut{KwOut}{Output}

    \KwIn{
    Production schedules variables: $\{Z_{s,t}\}_{s=1}^{96}\in \mathbb{R}^{72}$; Load process: $\{l_{s,t}\}_{s=1}^{96}$; Time step $s$ is 15-minute frequency, $t=1,2,\ldots,T$
    }
    \KwOut{
    Training Set: $\mathcal{S}^{train} = \{(X_{s,t}, y_{s,t+h})\}_{s}^{s=96}, h=1,2$, First 7-month sample $t$\\
    Test Set: $\mathcal{S}^{test} = \{(X_{s,t}, y_{s,t+h})\}_{s}^{s=96}, h=1,2$, Last 5-month sample $t$
    }

    Aggregate $\{Z_{s,t}\}_{s=1}^{96}\in \mathbb{R}^{72}$ into 3 groups, $\{x^{1}_{s,t}, x^{2}_{s,t}, x^{3}_{s,t}\}_{s=1}^{96}$, $t=1,2,\ldots,T$\;
    
    Predictor Variables: $\{X_{s,t}\}_{s=1}^{96} = \{x^{1}_{s,t}, x^{2}_{s,t}, x^{3}_{s,t}, l_{s,t}\}_{s=1}^{96}$, $t=1,2,\ldots,T$\;
    
    Forecast Load: $\{y_{p,t}\}_{p=1}^{192}$, $t=1,2,\ldots,T$\;
    
    Sample Pairs: $\{(X_{s,t}, y_{s,t+h})\}_{s}^{s=96}, h=1,2$, $t=1,2,\ldots,T$\;
    
    Split sample pairs into a training set (first 7-month) and test set (last 5-month)

    \caption{Forecasting Sample Construction}\label{algo:sample_construct}
\end{algorithm}

\subsection{Neural Network Models}
\label{sec:lstm}

Nonlinearity has been the ignored truth in statistics for a long time due to the interpretability of many nonlinear models. However, on the other hand, interpretation of the linear parametric models can sometimes be misleading when models are deviating too far away from the true model, likely caused by the poor flexibility of parametric models. To capture the nonlinear dependency of the load process, we exploit the power of machine learning techniques, especially Recurrent Neural Networks (RNN).

RNN is essentially the modified Neural Network (NN) architecture with a hidden state depending on the previous states and present information designed for the sequential problem. In general, the backpropagation algorithm (\cite{rumelhart1985learning,rumelhart1988learning}) is used to solve RNN problems. However, the RNN suffers from gradient-vanishing or gradient-exploding as the network. In turn, the network fails to capture long-memory effects(\cite{bengio1994learning}). \cite{hochreiter1997long} proposed a new design of recurrent neural unit called Long-short term memory unit (LSTM) that has "cell state" and "forget gate" which memorize and forget certain information, respectively.
Gated Recurrent Unit (GRU) provided by \cite{cho2014properties}) is one of the many variations of LSTM that aims to capture nonlinear temporal dependency but without the memory gate. GRU has performed well, even better than LSTM in some cases. We will describe details of LSTM and empirical forecast results.

\subsubsection{Long-Short Term Memory Unit}

\begin{figure}[!htb]
  \centering
  \begin{minipage}{1\linewidth}
    \includegraphics[width=\textwidth]{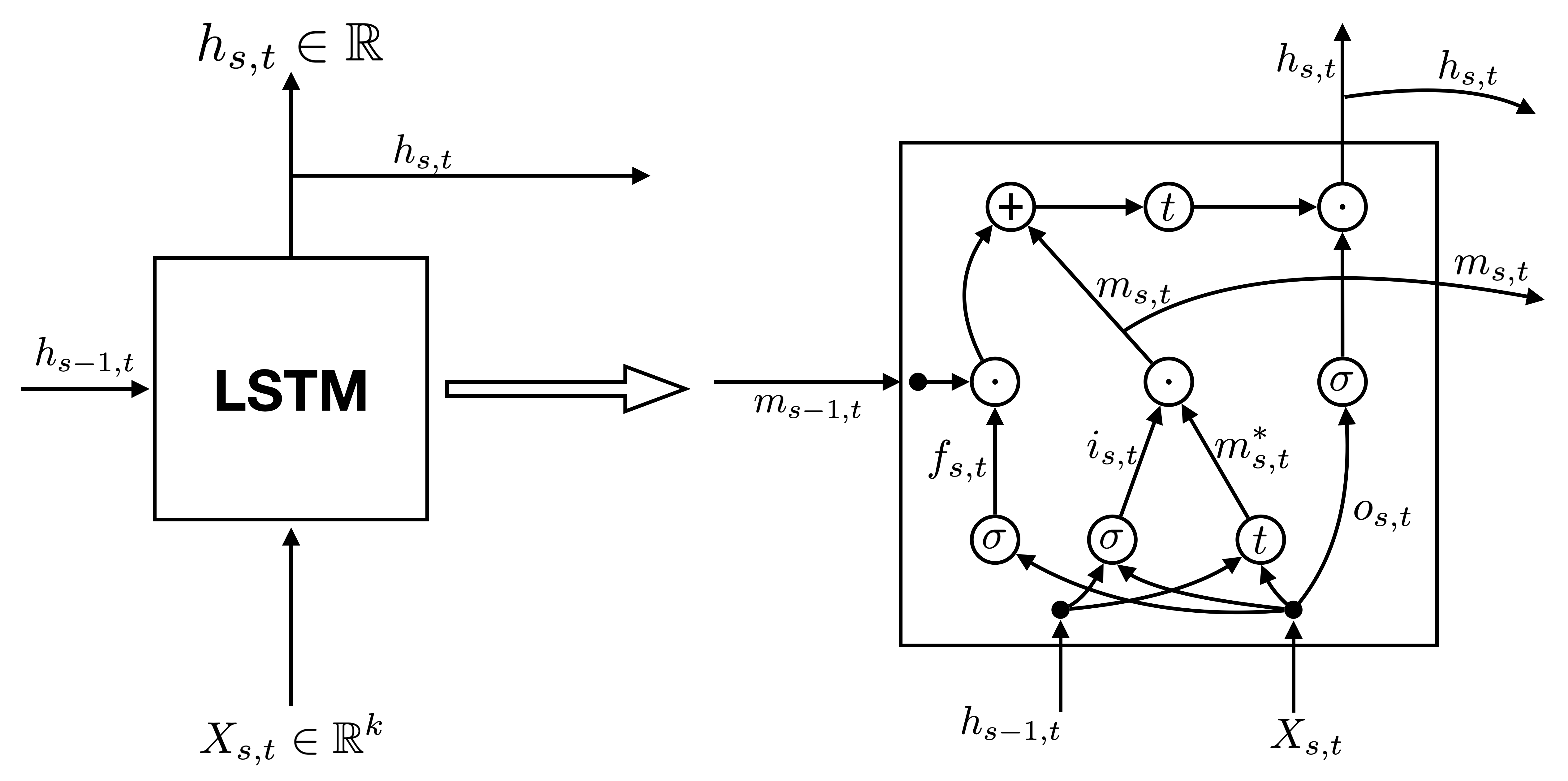}
  \end{minipage}
  \caption{
    \textbf{Architecture of LSTM}
  \newline 
  \small
The figure shows the architecture of a unit in the LSTM model. The left part shows the outside structure, where the input vector $X_{s,t}\in\mathbb{R}^k$ on the bottom and the output vector $h_{s,t}\in\mathbb{R}$ on the top, and the hidden state vector $h_{s-1,t}\in\mathbb{R}$ as a feedback input. The right part is the detailed structure inside the LSTM unit.
  }\label{fig:lstm_struct}
\end{figure}

LSTM models can be constructed in various ways depending on the number of layers and the number of neurons(units) of each layer of that network. On the other hand, there are only very limited versions of the LSTM unit. Here we employ only one layer of the most common LSTM unit finished with a dense layer. Each LSTM unit is shown in Figure \ref{fig:lstm_struct} and describe it as follows.

One LSTM unit has input of ${X_{s,t}}\in \mathbb{R}^{k}$ and output of $h_{s,t}\in \mathbb{R}$, for each time step $s$, and $t=1,2,\ldots,T$. The LSTM unit consists of 4 components called \enquote{gates}. The first gate is the \enquote{input gate} that has its output $i_{s,t}\in\mathbb{R}$ defined in Equation \eqref{eq:lstm_input} as a function of the the input vector $X_{s,t}$, the output from last step (memory), $h_{s-1,t}\in\mathbb{R}$, and the intercepts $\alpha_i\in\mathbb{R}$. Coefficients $\gamma^{s,i}\in \mathbb{R}^{k}$ and $\theta^{(s,i)}\in \mathbb{R}$ are estimated in each time step $s$. Note that $k$ represents the dimensionality of each sample at each time step. In our case, each sample is constructed with the lagged periods load value, 3 groups of lagged production schedules values, hence $k=4$. 

\begin{equation}\label{eq:lstm_input}
  i_{s,t} = N^{\sigma}\left(X_{s,t}^\top \gamma^{(s,i)} + h_{s-1,t}\theta^{(s,i)} + \alpha^{(s,i)}\right)
\end{equation}

The neural network activation function sigmoid function $N^{\sigma}$ and tanh function $N^{tanh}$ are element-wise operators defined as Equation \eqref{eq:lstm_acti}, respectively

\begin{align}\label{eq:lstm_acti}
  \begin{split}
    N^{\sigma}\left(x\right) &= \frac{1}{1+e^{-x}}\\
    N^{tanh}\left(x\right) &= \frac{e^x-e^{-x}}{e^x+e^{-x}}
  \end{split}
\end{align}

Then the forget gate is defined in Equation \eqref{eq:lstam_forget} which has the same structure as the input gate. This gate controls the amount of information that should be forgotten from the input $X_{s,t}$, previous output $h_{s-1,t}$.

\begin{equation}\label{eq:lstam_forget}
  f_{s,t} = N^{\sigma}\left(X_{s,t}^\top \gamma^{(s,f)} + h_{s-1,t}\theta^{(s,f)} + \alpha^{(s,f)}\right)
\end{equation}

The memory gate is defined as Equation \eqref{eq:lstm_memory}, where two parts of information are added up. The first part $f_{s,t}\odot m_{s-1,t}$ decides how much information should be kept from memory of last step $m_{s-1,t}$, while the second part is the new memory that is determined by the input $X_{s,t}$ and the output $h_{s-1,t}$ from previous step and the input $i_{s,t}$. The $\odot$ represents the element-wise product.

\begin{equation}\label{eq:lstm_memory}
    m_{s,t} = f_{s,t}\cdot m_{s-1,t} + i_{s,t}\cdot N^{tanh}\left(X_{s,t}^\top \gamma^{(s,m)} + h_{s-1,t} \theta^{(s,m)} + \alpha^{(s,m)}\right)
\end{equation}

The $o_{s,t}$ is $N^{\sigma}$ function of linear combination of input $X_{s,t}$, output from last period $h_{s-1,t}$. The output vector $h_{s,t}$ is determined by $o_{s,t}$ and $m_{s,t}$.

\begin{align}
  \begin{split}
    o_{s,t} &= N^{\sigma}\left(X_{s,t}^\top \gamma^{(s,o)} + h_{s-1,t}\theta^{(s,o)} + \alpha^{(s,o)}\right)\\
    h_{s,t} &= o_{s,t} \cdot N^{tanh}\left(m_{s,t}\right)
  \end{split}
\end{align}

For each sample, the LSTM unit will produce output vector $H_{t}=\left(h_{1,t}, h_{2,t}, \ldots, h_{S,t}\right)^\top$, $h_{s,t}\in \mathbb{R}$ for $s=1,2,\ldots,S$, and $S=96$. Finally, a dense layer is employed to transform the $H_{t}$ to $y_{t+2}$,

\begin{equation}
    y_{t+2} = D \cdot H_{t},
\end{equation}

where $D$ is a matrix, $D\in \mathbb{R}^{P\times S}$.

Each gate has three parameter vectors to estimated. Take the input gate for example,

\begin{align}
    \begin{split}
        \gamma^{i} &= \left(\gamma^{1,i}, \gamma^{2,i}, \ldots, \gamma^{S,i}\right)^\top\\
        \theta^{i} &= \left(\theta^{1,i}, \theta^{2,i}, \ldots, \theta^{S,i}\right)^\top\\
        \alpha^{i} &= \left(\alpha^{1,i}, \alpha^{2,i}, \ldots, \alpha^{S,i}\right)^\top\\
    \end{split}
\end{align}

Similarly, the forget gate has $\gamma^{f}$, $\theta^{f}$, $\alpha^{f}$; the memory gate has $\gamma^{m}$, $\theta^{m}$, $\alpha^{m}$; and the output gate has $\gamma^{o}$, $\theta^{o}$, $\alpha^{o}$. Therefore, the parameter matrix can be written as,

\begin{align}
    \begin{split}
    \Gamma &= (\gamma^{i}, \gamma^{f}, \gamma^{m}, \gamma^{o})\in \mathbb{R}^{S\times k*4}\\
    \Theta &= (\theta^{i}, \theta^{f}, \theta^{m}, \theta^{o})\in \mathbb{R}^{S\times 4}\\
    \Lambda &= (\alpha^{i}, \alpha^{f}, \alpha^{m}, \alpha^{o})\in \mathbb{R}^{S\times 4}\\
    \end{split}
\end{align}

The number of parameters estimated in the LSTM model heavily depends on the number of time steps, $S$. We employ a 1-day lagged sample of a 15-min sampling frequency, i.e., $S=96$.

\subsubsection{Gated Recurrent Unit}

\begin{figure}[!htb]
  \centering
  \begin{minipage}{1\linewidth}
    \includegraphics[width=\textwidth]{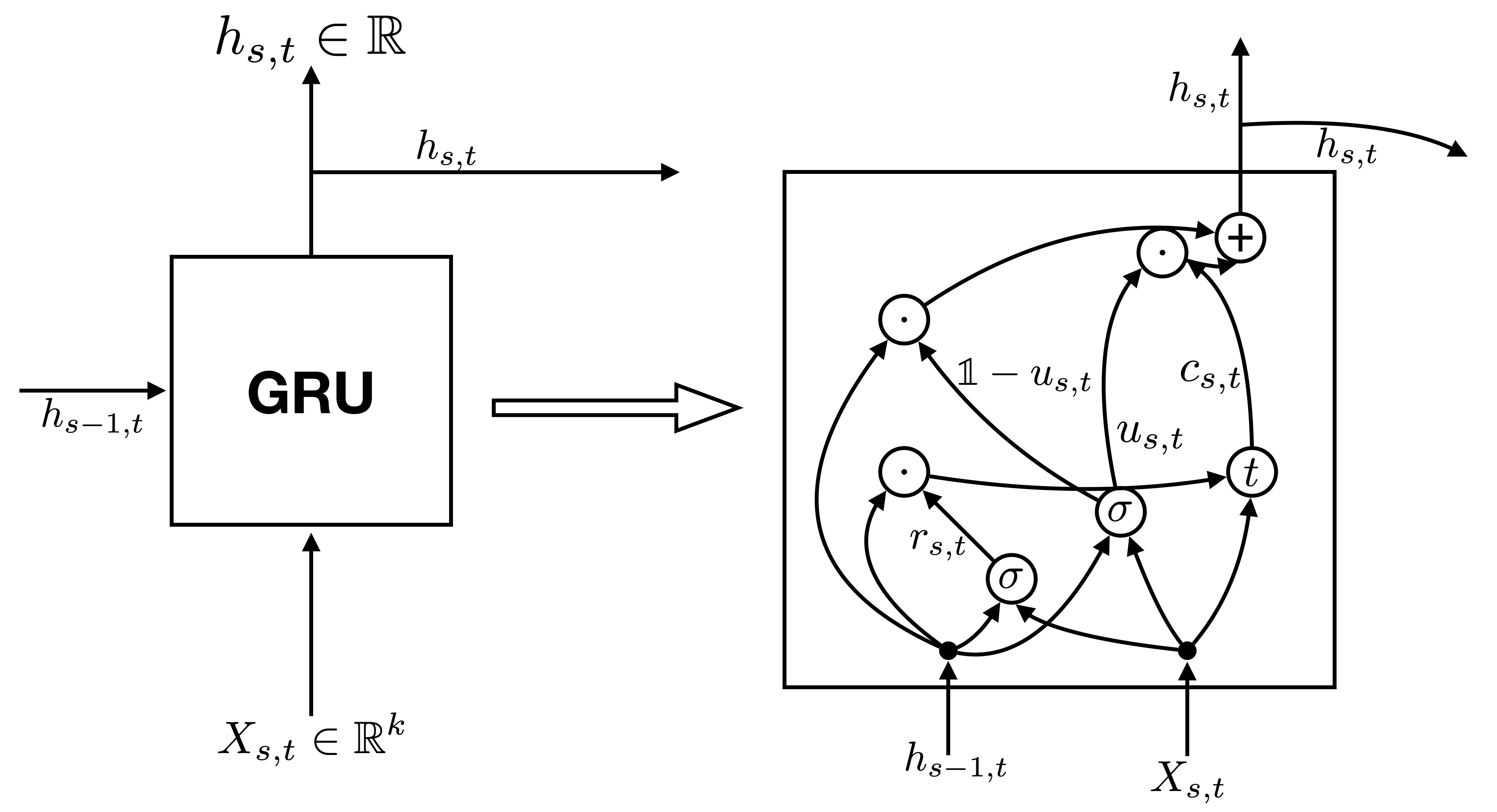}
  \end{minipage}
  \caption{
    \textbf{Architecture of GRU}
  \newline 
  \small
The figure shows the architecture of a unit in the GRU model. The left part shows the outside structure, where the input vector $X_{s,t}\in\mathbb{R}^k$ on the bottom and the output vector $h_{s,t}\in\mathbb{R}$ on the top, and the hidden state vector $h_{s-1,t}\in\mathbb{R}$ as a feedback input. The right part is the detailed structure inside the GRU unit.
  }\label{fig:gru_struct}
\end{figure}

We can see the graphic representation of GRU architecture clearly from Figure \ref{fig:gru_struct}. Instead of having memory gate and forget gate, GRU has a candidate activation $c_{s,t}$. And the output of GRU $h_{s,t}\in\mathbb{R}^{d}$ is linear interpolation of $c_{s,t}$ and $h_{s-1,t}$, as shown in Equation \eqref{eq:gru_output}, the update weight is determined by $u_{s,t}$ in Equation \eqref{eq:gru_update}.

Similar to LSTM, The input gate for GRU defined by Equation \eqref{eq:gru_reset} is also linear combination of input variable $X_{s,t}\in\mathbb{R}^k$ and previous output $h_{s-1,t}$,

\begin{equation}\label{eq:gru_reset}
  i_{s,t} = N^{\sigma}\left(X_{s,t}^\top \gamma^{(s,i)} + h_{s-1,t}\theta^{(s,i)} + \alpha^{(s,i)}\right)
\end{equation}

Differently, the candidate activation $c_{s,t}$ defined in Equation \eqref{eq:gru_candidate} is a function of $X_{s,t}$ and $i_{s,t}\cdot h_{s-1,t}$. The candidate activation $c_{s,t}$ determines how much information will should be kept or reduced from the input and previous result via the $N^{tanh}$ neural network layer.

\begin{equation}\label{eq:gru_candidate}
    c_{s,t} = N^{tanh} \left(X_{s,t}^\top \gamma^{(s,c)} + \left(i_{s,t}\theta^{(s,c)}\cdot h_{s-1,t}\right) + \alpha^{(s,c)}\right)
\end{equation}

The interpolation weighting vector $u_{s,t}\in\mathbb{R}$ defined in Equation \eqref{eq:gru_update} controls the degree of updating from candidate activation $c_{s,t}$.

\begin{equation}\label{eq:gru_update}
    u_{s,t} = N^{\sigma}\left(X_{s,t}^\top \gamma^{(s,u)} + h_{s-1,t}\theta^{(s,u)} + \alpha^{(s,u)}\right)
\end{equation}

The output gate keeps part of the information from the previous output and part of the information from candidate activation $c_{s,t}$.

\begin{equation}\label{eq:gru_output}
    h_{s,t} = \left(\mathds{1}-u_{s,t}\right)\cdot h_{s-1,t} + u_{s,t}\cdot c_{s,t}
\end{equation}

For each sample, the GRU unit output vector is $H_{t}=\left(h_{1,t}, h_{2,t}, \ldots, h_{S,t}\right)^\top$, $h_{s,t}\in \mathbb{R}$ for $s=1,2,\ldots,S$, and $S=96$. Same to LSTM model, a dense layer with linear transform $D\in \mathbb{R}^{P\times S}$ is employed to transform the $H_{t}$ to $y_{t+2}$ finally,

\begin{equation}
    y_{t+2} = D \cdot H_{t},
\end{equation}

As a GRU unit has one less gate compared to a LSTM unit, the number of parameters to be estimated in the GRU unit is less than in the GRU unit,

\begin{align}
    \begin{split}
    \Gamma &= (\gamma^{i}, \gamma^{c}, \gamma^{u})\in \mathbb{R}^{S\times k*3}\\
    \Theta &= (\theta^{i}, \theta^{c}, \theta^{u})\in \mathbb{R}^{S\times 3}\\
    \Lambda &= (\alpha^{i}, \alpha^{c}, \alpha^{u})\in \mathbb{R}^{S\times 3}\\
    \end{split}
\end{align}

\subsubsection{Parameters Estimation}

Due to the flexibility and complexity of LSTM/GRU, a closed-form solution for estimating parameters is not possible. In general, all parameters are estimated by the backpropagation (BPP) algorithm described as follows.

Let $\hat{\Psi}$ be the estimated coefficients matrix in a complete constructed LSTM/GRU model which can be formulated in a general form stated as Equation \eqref{eq:nn_exp}. The forecast $y_t$ at step $t$ is determined by function $g(\cdot)$ which is the nonlinear neural network with forcastors $X_{s,t}$.

\begin{equation}\label{eq:nn_exp}
  y_t = g\left(\Psi;X_{s,t},h_{s-1,t}\right) + \varepsilon_t
\end{equation}

In BPP, all parameters are estimated by iteration shown as Equation \eqref{eq:bpp} with random initial values. $\delta$ is a prespecified constant that determines the learning rate of such a neural network, and $\nabla$ is the gradient operation.

\begin{equation}\label{eq:bpp}
  \hat{\Psi}^{(j+1)} = \hat{\Psi}^{(j)} + \delta\cdot\nabla g\left(\hat{\Psi}^{(j)};X_{s,t}, h_{s-1,t}\right)\left[h_{s,t} - g\left(\hat{\Psi}^{(j)};X_{s,t}, h_{s-1,t}\right)\right],i=1,2,...
\end{equation}

\subsection{FASTEC on Regime Switching}

\subsubsection{FActorisable Sparse Tail Event Curves}

In the following, we present a functional approach to model the load process of chemical production. This method models the short-term electricity consumption $Y_t$ around a local mean, trend, or seasonality. The proposed method is based on a sequence of steps: first, we estimate a functional factor model at each production level using expectiles. Expectiles can be referred to as conditional moments at a specific level of the distribution. The functional approach employs $B$-splines as regressors and, hence, cannot be used directly for forecasting purposes. By employing an asymmetric loss function, it is possible to estimate the model coefficients conditional on being in a specific region of the distribution, such that the approximation of the entire probability distribution of the stochastic variation is achieved without imposing the type of distribution. Within the estimation procedure, iteratively singular value decomposition is procured in the coefficient estimation step. Left singular-vectors and the singular values give us time-invariant factors that capture the within-day dynamic of our functional data, and time-varying loadings, the right-hand singular vectors, can be deployed in a time series regression model. Then, a vector autoregressive time series model is estimated on the time-varying loadings of the first $r$ most relevant factors to capture any short term variation that might occur in the data.

Expectile regressions have become popular functional data analysis (FDA) techniques (\citet{Chao:2015b}, \citet{Schnabel:2009}, \citet{Schulz:2014}, \citet{Sobotka:2013}, \citet{Tran:2018}, \citet{Ramsay:2007}, \citet{burdejova2019dynamic}). The multivariate expectile regression is done using the FActorisable Sparse Tail Event Curves (FASTEC) methodology, a recent technique involving generalized quantiles in a multivariate regression setup. It incorporates quantile curves \citep{Chao:2015b} or expectile curves \citep{Huang:2016} in a linear model that represents functional variables as linear combinations of $r$ $B$-splines. This approach is in spirit analog to the Karhunen-Lo\`{e}ve expansion or the Fourier series that also employ linear combinations of orthogonal functions. An approximation of each load curve is then given by the sum of the first $r$ of these functions.

In a general setup, suppose  $\{{\bf X}_{s}\}_{s=1}^{S},  \in \mathbb{R}^{k},\{{\bf Y}_{s}\}_{s=1}^{S} \in \mathbb{R}^q$ i.i.d.; The multivariate linear model for a $\tau$-expectile curve of the daily curve $Y_{t}$, consisting of $S$ intra-day observations, with $t=1,...,T$, and the level $\tau \in (0, 1)$, is given by

\begin{align}
    Y_t &= e_t(\tau|{\bf X}_s) + U_{st, \tau} = {\bf X}_s^\top \Gamma_{\ast t}(\tau) + U_{st, \tau} \label{eqno:01}
\end{align}

where $\Gamma_{\ast t}(\tau) \in \mathbb{R}^{k}$ is the coefficient matrix for day $t$ at $\tau$-level and $e_t(\tau|{\bf X}_s)$ is the conditional expectile of $Y_t$ given ${\bf X}_s$ and $U_{st, \tau}$ is $\tau$-level white noise. $k$ is the number linear combinations of B-splines required to represent the $Y_t$ and is chosen as $\floor*{\sqrt{n}}$.

The multivariate expectile regression formulation with penalised loss is given by

\begin{align}
{\Gamma}_{\lambda}(\tau)= \argmin_{\Gamma\in\mathbb{R}^{k\times q}}\left\{ (TS)^{-1}\sum_{s=1}^S\sum_{t=1}^T\rho_\tau\left(Y_{st} - {\bf X}^\top_{s}\Gamma_{\ast t}\right) + \lambda\|\Gamma\|_{\ast}\right\},\label{eq:03}
\end{align}

where $\|\Gamma\|_\ast = \sum_{t=1}^{\min(k,q)}\sigma_t(\Gamma)$ is the nuclear norm of $\Gamma$, ${\bf X}_{s}$ is a $(S \times k)$-matrix, $\Gamma$ is the coefficient matrix and $\lambda$ the penalisation parameter. Please note, that $\Gamma_\lambda(\tau) = \Gamma$ depends on the tuning parameter $\lambda$ and the $\tau$-level. To ensure efficiency and convergence of the algorithm optimal selection of the tuning parameter $\lambda$ is of the essence. The $n$-fold cross-validation method gives a consistent estimate of the tuning parameter \citep{Haerdle:2012:SPM}. In our case, we are required to forecast the load process 2-day ahead, the FASTEC method is applied twice each time to forecast a 2-day result.

The decision to which expectile level the current data belongs and, hence, where the forecast of the load will be, is entirely data-driven and is evaluated by a change-point test. The change-point detection is based on a $t$-test that captures the mean distance of curve $Y_t$ to all mean levels estimated from the Gaussian Mixture Model described later in Sec.\ref{sec:gmm}, where the load process is estimated in a 15-minute frequency. Thus, the expected distance to one of the levels is expected to be zero. If this is the case, the test-statistic is small, and the null hypothesis of having the same level cannot be rejected.

\subsubsection{Gaussian Mixture Model}\label{sec:gmm}

\begin{figure}[!htb]
  \centering
  \begin{minipage}{1\linewidth}
    \includegraphics[width=\textwidth]{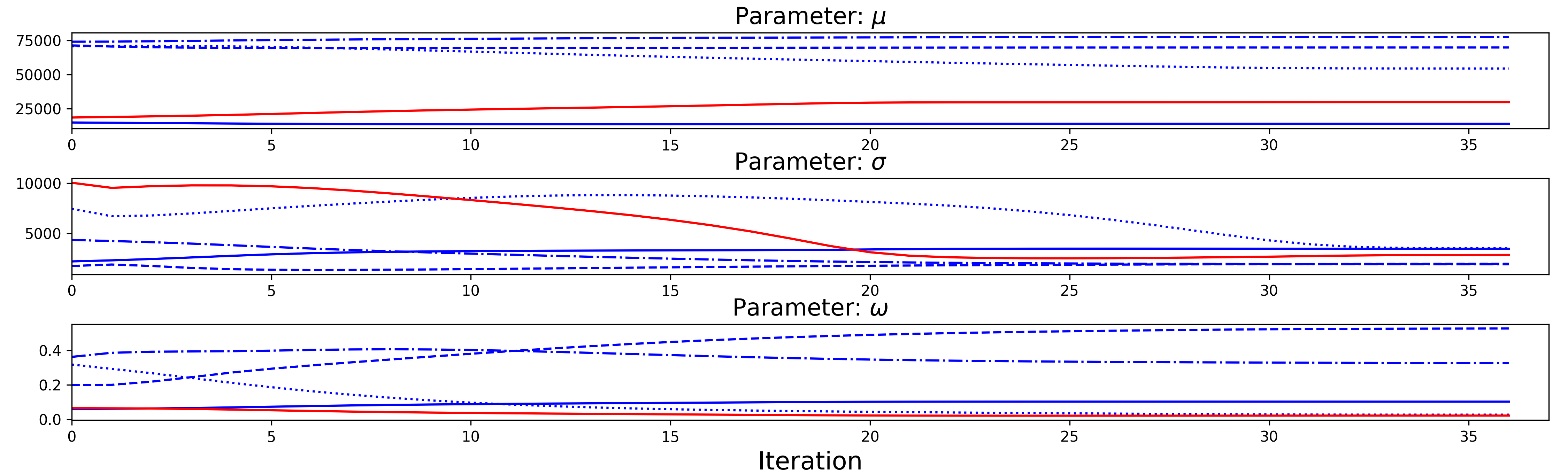}
  \end{minipage}
  \caption{
    \textbf{Convergence of EM algorithm on load process}
  \newline 
  \small
  The figure shows how the parameters converge with the EM algorithm.
  }\label{fig:em_converge}
\end{figure}

Motivated by Figure \ref{fig:load_kde}, the distribution of load process points can be modeled as Gaussian Mixture Model (GMM) (\cite{mclachlan2004finite}, \cite{10.1007/978-3-642-41398-8_15}).
Here, we model the GMM using the first 10,000 points, i.e $D=10,000$.
GMM assumes that the probability density function of the load process $\{y_d\}_{d=1}^{D}$ in a 15-minute frequency can be written as a one-dimensional linear combination of $k$ Gaussian density functions as follows.

\begin{equation}\label{eq:gmm}
    P(y| \alpha, \mu, \sigma^2) = \sum^{K}_{k=1}\alpha_{k}\varphi(y | \mu_k, \sigma_k^2)
\end{equation}

Where $\Theta_k = \left(\alpha_k, \mu_k, \sigma_k^2 \right)^\top$ are the weight, mean, and variance of the $k_{th}$ normal density function, respectively.
Under the restriction that $\alpha_k\in(0,1),k=1,...,K$, and $\sum^{K}_{k=1}\alpha_k = 1$.

In this article, the goal of GMM modeling is to detect the mean values for different states in the load process. Thus we ignore the temporal dependency of the load process. The estimation procedure is carried by the classic Expectation-Maximization (EM) algorithm provide by \cite{dempster1977maximum} and then extended by \cite{ruud1991extensions}.

To estimate the parameters in Equation \eqref{eq:gmm}, we assume that there are $K=5$ machinery work states, the latent variable $S_{t}^{(k)}$ represents that machine works on the $k_{th}$ status at time $t$. $S_t^{(k)}$ is an indicator function that equals 1 with probability of $\alpha_k$. Conditioning on the working status, the the probability density function is Gaussian, i.e $\varphi(y_t | S_t^{(k)}, \mu_k, \sigma^2_k)$. Then, the log-likelihood function under the assumptions can be written as:

\begin{align}\label{eq:em_loglikelihood}
  \begin{split}
    \log P(y, S^{(k)} | \Theta) &= \log\left(P(y | S^{(k)}, \Theta) \cdot P(S^{(k)}| \Theta) \right) \\
          &= \log \left( \prod^{K}_{k=1}\alpha_k^{n_k}\cdot\prod^{T}_{t=1} \varphi(y_t | \mu_k, \sigma_k)^{S_t^{(k)}} \right)\\
          &= \sum^{K}_{k=1}\left\{n_k \log\alpha_k + \sum^{T}_{t=1}S_t^{(k)} \left[\log\frac{1}{\sqrt{2\pi}} - \log\sigma_k - \frac{(y_t -\mu_k)^2}{2\sigma_k}\right] \right\}
  \end{split}
\end{align}

Where $n_k=\sum^{T}_{t=1}S^{(k)}_t$ is the number of points that belong to the $k_{th}$ status.

The E-step of the EM algorithm is taking the expectation of the likelihood function in Equation \eqref{eq:em_loglikelihood} named $\mathbb{Q}$ function.

\begin{align}
  \begin{split}
    \mathbb{Q}\left(\Theta, \Theta^{(i-1)}\right) &= \operatorname{E}\giventhat[\Big]{\log P(y, S^{(k)} | \Theta)}{y, \Theta^{(i-1)}}\\
    &= \sum^{K}_{k=1} \sum^{T}_{t=1}\operatorname{E}\left(S_t^{(k)}|y, \Theta^{(i-1)}\right)\log\alpha_k \\
&+ \sum^{K}_{k=1}\sum^{T}_{t=1}\operatorname{E}\left(S_t^{(k)}|y, \Theta^{(i-1)}\right) \left[\log\frac{1}{\sqrt{2\pi}} - \log\sigma_k - \frac{(y_t -\mu_k)^2}{2\sigma_k}\right]
  \end{split}
\end{align}

The conditional expectation of working status at time $t$ is estimated as follows.

\begin{equation}
  \operatorname{E}\left(S_t^{(k)}|y, \Theta^{(i-1)}\right)=\frac{\alpha_k\varphi(y_t| \Theta_k^{(i-1)})}{\sum^{K}_{k=1}\alpha_k\varphi(y_t| \Theta_k^{(i-1)})}
\end{equation}

Then the M-step of the EM algorithm is the maximization of $\mathbb{Q}$ function with iteration,

\begin{equation}
  \Theta^{i} = \argmax_{\Theta}\mathbb{Q}\left( \Theta, \Theta^{(i-1)}\right)
\end{equation}

As shown in Figure \ref{fig:em_converge}, the EM algorithm converges quickly. With the estimated states mean $\mu$, the change-point test of FASTEC can then be performed.

\begin{algorithm}[!htb]
    \SetKwFunction{fastec}{FASTEC}
    \SetKwFunction{gmm}{Gaussian Mixture Model}
    \SetKwInOut{KwIn}{Input}
    \SetKwInOut{KwOut}{Output}
    \SetKwProg{Fn}{Function}{:}{End Function}

    \KwIn{
    Training Sample $\mathcal{S}^{train} = \{(X_{s,t}, y_{s,t+h})\}_{s}^{s=96}, h=1,2$ from Algo. \ref{algo:sample_construct}
    }
    \KwOut{
    Estimated coefficient matrix: $\Gamma_{*t}(\tau)$
    }
    
	Estimate conditional moments: certain expectiles as cluster centers\;
    \Indp
	Perform the Gaussian Mixture Model with load process in training sample $\{y_{s,t}\}_{s=1}^{96}$ to obtain $\mu$ and $\sigma$ at each level\;
	\Indm
	Determine affiliation based on $t$-statistic with estimated $\mu$ and $\sigma$\;
	Perform FASTEC (training) on $h=1,2$ separately:\\
		\Indp 
		Select periods at conditional moments with 96 observations $\{y_{s,t+h}^\tau\}_{s=1}^{96}$\;
		Bootstrap to get larger sample $y_{s,t+h}^{\tau\ast}$\;
		Perform FASTEC on each conditional moment sample\;
		Predict conditional moments $\widehat{y}_{s,t+h}^{\tau\ast}$\;
    \Indm
    Estimate the weights vector $\widehat{\gamma}$ for half of each sample $\{\widehat{y}_{s,t+h}^{\tau\ast}, x_{s,t}^{1}, x_{s,t}^{2}, x_{s,t}^{3}\}_{s=1+96*h}^{96*h}$ s.t. MAE is minimised\;
    \caption{Algorithm of FASTEC Method with Regime Switching}\label{algo:fastec}
\end{algorithm}

We summarize the FASTEC method with regime switching as the following algorithm representation in Algorithm \ref{algo:fastec}.

\subsection{Performance Evaluation}

To have a more robust comparison, we employ five metrics described as follows to measure our forecast results.

\begin{figure}[!htb]
  \centering
  \begin{minipage}{1\linewidth}
    \includegraphics[width=\textwidth]{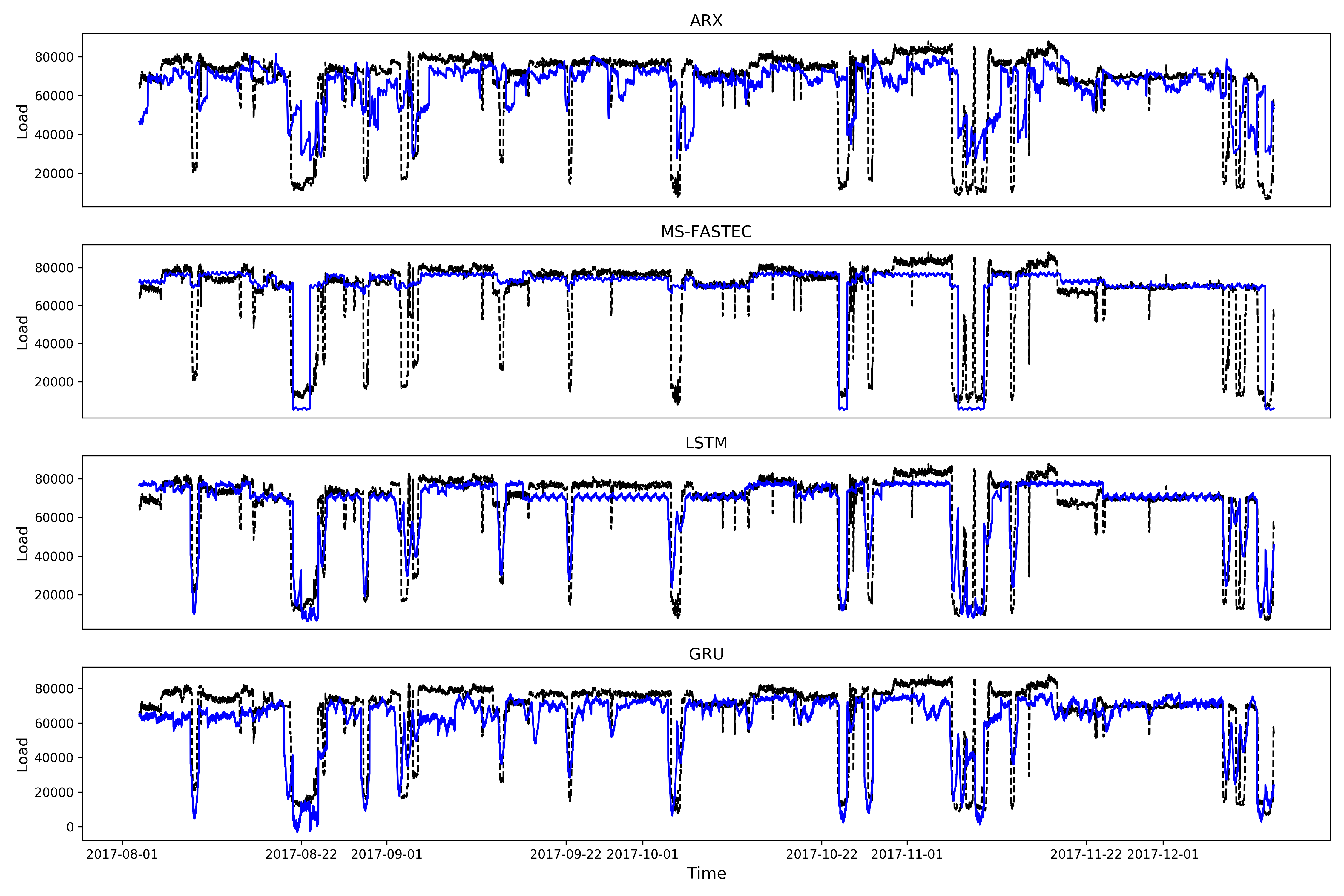}
  \end{minipage}
  \caption{
    \textbf{Out-of-Sample Forecasting Comparison}
  \newline 
  \small
The figure shows the out-of-sample forecasting results (\textcolor{blue}{blue line}) from all the models compare against the true load process (black line).
  }\label{fig:full_sample_compare}
\end{figure}

The Mean Absolute Error is defined as $MAE = \frac{1}{T}\sum_{t=1}^T\left|{y_t-\widehat{y}_t}\right|$, where $T$ is the out-of-sample points. $MAE$ measures the average absolute distance between the true values, $y_t$, and forecasts, $\widehat{y}_t$. Unfortunately, the naive approach (taking value from previous period) would take advantage of $MAE$ metrics if the process is evolving mildly. Using the Mean Absolute Scaled Error, $MASE = \frac{1}{T}\sum_{t=1}^T l^{-1} |y_t-\widehat{y}_t|$, overcomes such bias by adjusting $MAE$ with $l = \frac{1}{T-1}\sum_{s=2}^{T}|y_s-y_{s-1}|$ that accounts the average changing intensity between adjacent periods. To relieve the influence from scale, we also measure with the Mean Absolute Percentage Error $MAPE = \frac{1}{T}\sum_{t=1}^T\left|(y_t-\widehat{y}_t)/y_t\right|$.

The Mean Squared Error $MSE = T^{-\frac{1}{2}}\sqrt{\sum_{t=1}^T (\widehat{y}_t-y_t)^2}$ measures the average Euclidean distance between true values and forecasts. We employ 3 metrics that are essentially $MSE$ adjusted by 3 different factors, $Q$, namely the Root Mean Squared Error $RMSE = T^{-\frac{1}{2}}Q^{-1}\sqrt{\sum_{t=1}^T (\widehat{y}_t-y_t)^2}$. The first one $nrRMSE$ is adjusted by difference between the largest and smallest values, $Q=Q_{nr}=\{max(y_t)-min(y_t)\}$, the second one $nmRMSE$ is adjusted by the mean value $Q=Q_{nm}=T^{-1}\sum^{T}_{t=1}y_t$, and the thrid one $niqrRMSE$ is ajusted by empirical quantile difference, $Q=Q_{niqr}=\{q(y_t, 0.75)-q(y_t, 0.25)\}$.

Then Diebold-Mariano test (DM) provided by \cite{diebold2002comparing} is employed to compare forecasting results from different models. D-M test compares the forecasting errors from two models and gives the statistics of whether one error process is significantly smaller or larger than the other one. Here we compare LSTM against GRU, MS-FASTEC, and ARX models using quadratic loss, i.e the loss differential $d^{LSTM,m}_{t} = (y^{LSTM}_{t}-y_t)^2 - (y^{m}_{t}-y_t)^2$
where $m\in\left\{GRU, MS\mbox{-}FASTEC,ARX\right\}$, and $y_t$ is the true value at time $t$.
Then the sample mean loss differential is $\bar{d}_{t}^{LSTM,m} = T^{-1}\sum^{T}_{t=1}d^{LSTM,m}_{t}$. Under some assumption, the D-M test is formulated as follows:

\begin{equation}
  DM_{t}^{LSTM,m} = \frac{\bar{d}_{t}^{LSTM,m}}{\hat{\sigma}_{\bar{d}_{t}^{LSTM,m}}}
\end{equation}

Where $\hat{\sigma}_{\bar{d}_{t}^{LSTM,m}}$ is the consistent estimate of standard deviation of $\bar{d}_{t}^{LSTM,m}$.

\begin{table}[!htb]
  \small
  \setlength{\tabcolsep}{0pt}
  \begin{threeparttable}
    \caption{Out-of-sample 2-days ahead forecast performance evaluation}\label{tab:forecast_eval}
    \begin{tabular*}{\linewidth}{@{\extracolsep{\fill}}>{\itshape}l *{1}{S[table-format=1.4,table-number-alignment = center]}
    *{3}{S[table-format=3.5,table-number-alignment = center]}
    }
        \toprule\toprule
        
    \multicolumn{1}{c}{Measure} & \multicolumn{1}{c}{LSTM} & \multicolumn{1}{c}{GRU} & \multicolumn{1}{c}{MS\mbox{-}FASTEC} & \multicolumn{1}{c}{ARX} \\
      \hline
      MAPE                      & 0.211$^{\dagger}$        & 0.242                   & 0.332                                & 0.415                   \\
      MASE                      & 0.597$^{\dagger}$        & 0.848                   & 0.672                                & 1.040                   \\
      nRMSE                     & 0.142$^{\dagger}$        & 0.175                   & 0.197                                & 0.220                   \\
      niqRMSE                   & 1.323$^{\dagger}$        & 1.628                   & 1.831                                & 2.052                   \\
      nmRMSE                    & 0.173$^{\dagger}$        & 0.212                   & 0.239                                & 0.268                   \\
      D-M$^{\ddagger}$          &                          & -19.138$^{***}$          & -21.556$^{***}$                       & -31.297$^{***}$          \\
        
          \bottomrule\bottomrule
        
    \end{tabular*}
  \begin{tablenotes}[flushleft]
    \setlength\labelsep{0pt}
  \linespread{1}\small
  \item 
  $\dagger$: Model with the highest accuracy compared with other models under each metric.
  \item
  $\ddagger$: DM test with quadratic loss. GRU, MS-FASTEC, and ARX models against the LSTM model. A significant negative testing result implies that the LSTM model outperforms the other models.
  $^{***}$: Signficant at 1\% level.
  \end{tablenotes}
  \end{threeparttable}
\end{table}

Table \ref{tab:forecast_eval} shows the metrics of different models, and the LSTM model outperforms parametric models and even GRU consistently. This result strongly suggests that using a suitable neural network model can perform much better when the nonlinearity property is captured.

\begin{figure}[!htb]
  \centering
  \begin{minipage}{1\linewidth}
    \includegraphics[width=\textwidth]{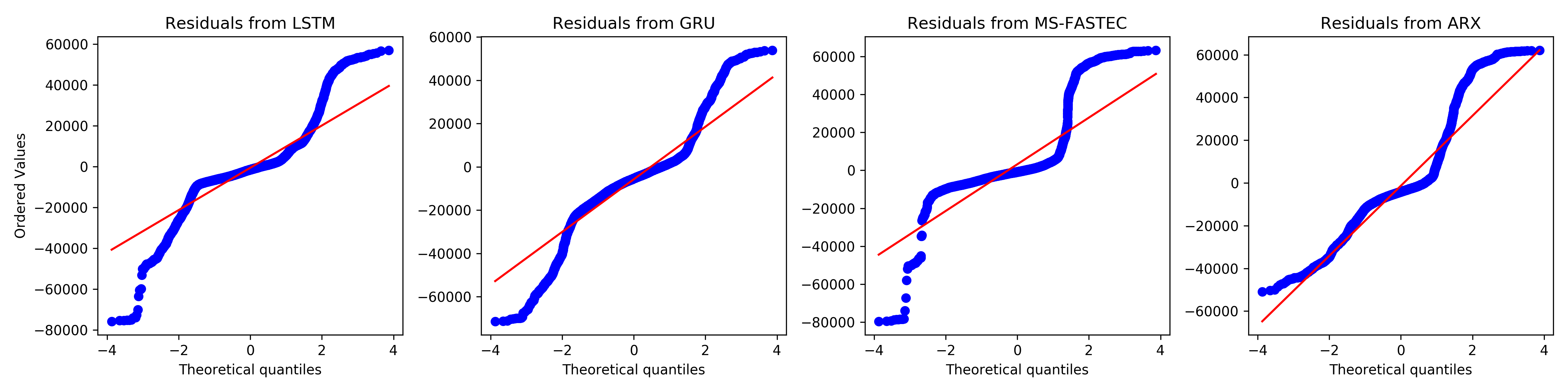}
  \end{minipage}
  \caption{
    \textbf{Distribution Residuals from Each Forecasting model}
  \newline 
  \small
The Q-Q plot shows the distribution of residuals from each model against the standard normal distribution. Models from left to right plots are LSTM, GRU, MS-FASTEC, ARX.
  }\label{fig:residual_qqplot}
\end{figure}

Even though the improvement from the LSTM model is significant, Figure \ref{fig:residual_qqplot} shows that none of the models could generate a forecasting error process that is normally distributed. One explanation of such forecasting results is that in our case, the load process is not surrendered to any clear pattern and is filled with a huge amount of jumps.

\section{Conclusion}\label{conclusion}

Short term load forecast is crucial for energy consumers, especially the big manufacturers, for the forecast is one of the key parts of their decision-making. Previously, most of the studies focus on modeling the load process from large-scale consumers, for example, a countrywide electricity consumption process that evolves with a seasonal or periodic pattern. Forecast on the small-scale consumer, such as a household or a set of machines, is a pragmatic problem for the industries while rarely investigating. Requesting by a chemical production company located in Germany, we are motivated to investigate the electric load process of a chemical production line that involves hard structures.

Despite the fact that the load process has a strong temporal dependency, which can be captured by autoregression models or long-memory models, one can see a huge amount of jumps in the process. Some of the jumps exist only for a short time, and some of them actually represent the load process transfer to another working status. The nonlinear dependence motivates us to implement state-of-the-art neural network models.

We first calibrate two of the popular deep neural network models, LSTM and GRU. The parameters are estimated by the standard backpropagation algorithm. To capture the tail events of the process, we further construct a functional data analysis based semi-parametric model, FASTEC, combined with the EM algorithm. For comparison, we use the ARX model as a benchmark.

Utilizing the one-year intraday data, all models forecast the electricity consumption in the next two days in 15-min resolution. We train all the models with the first 7-month data and conduct out-of-sample forecasting in the last 5-month. The out-of-sample forecast suggests that the LSTM model outperforms all other models by all metrics significantly evaluated by the D-M test. However, even the LSTM model can not harness the tail risks caused by jumps, which is one crucially problem for further researches.

\clearpage

\appendix

\section{An Appendix}
\label{sec:app1}


\bibliographystyle{jfe}
\bibliography{ref}

\begin{thebibliography}{41}
\expandafter\ifx\csname natexlab\endcsname\relax\def\natexlab#1{#1}\fi

\bibitem[{Amara et~al.(2017)Amara, Agbossou, Dub{\'{e}}, Kelouwani, Cardenas,
  and Bouchard}]{Amara2017}
Amara, F., Agbossou, K., Dub{\'{e}}, Y., Kelouwani, S., Cardenas, A., Bouchard,
  J., 2017. Household electricity demand forecasting using adaptive conditional
  density estimation. Energy And Buildings 156, 271--280.

\bibitem[{An et~al.(2013)An, Zhao, Wang, Shang, and Zhao}]{an2013using}
An, N., Zhao, W., Wang, J., Shang, D., Zhao, E., 2013. Using multi-output
  feedforward neural network with empirical mode decomposition based signal
  filtering for electricity demand forecasting. Energy 49, 279--288.

\bibitem[{Bengio et~al.(1994)Bengio, Simard, and Frasconi}]{bengio1994learning}
Bengio, Y., Simard, P., Frasconi, P., 1994. Learning long-term dependencies
  with gradient descent is difficult. Ieee Transactions On Neural Networks 5,
  157--166.

\bibitem[{Bianchi et~al.(2017)Bianchi, Maiorino, Kampffmeyer, Rizzi, and
  Jenssen}]{bianchi2017overview}
Bianchi, F.~M., Maiorino, E., Kampffmeyer, M.~C., Rizzi, A., Jenssen, R., 2017.
  An overview and comparative analysis of recurrent neural networks for short
  term load forecasting. Arxiv Preprint Arxiv:1705.04378 .

\bibitem[{Bianco et~al.(2009)Bianco, Manca, and
  Nardini}]{bianco2009electricity}
Bianco, V., Manca, O., Nardini, S., 2009. Electricity consumption forecasting
  in italy using linear regression models. Energy 34, 1413--1421.

\bibitem[{Burdejov{\'a} and H{\"a}rdle(2019)}]{burdejova2019dynamic}
Burdejov{\'a}, P., H{\"a}rdle, W.~K., 2019. Dynamic semi-parametric factor
  model for functional expectiles. Computational Statistics 34, 489--502.

\bibitem[{Castelli et~al.(2015)Castelli, Vanneschi, and Felice}]{Castelli_2015}
Castelli, M., Vanneschi, L., Felice, M.~D., 2015. Forecasting short-term
  electricity consumption using a semantics-based genetic programming
  framework: The south italy case. Energy Economics 47, 37--41.

\bibitem[{Chang et~al.(2011)Chang, Fan, and Lin}]{chang2011monthly}
Chang, P.-c., Fan, C.-y., Lin, J.-j., 2011. Monthly electricity demand
  forecasting based on a weighted evolving fuzzy neural network approach.
  International Journal Of Electrical Power \& Energy Systems 33, 17--27.

\bibitem[{Chao et~al.(2018)Chao, H{\"a}rdle, and Huang}]{Huang:2016}
Chao, S.-k., H{\"a}rdle, W.~K., Huang, C., 2018. {Multivariate factorizable
  expectile regression with application to fMRI data}. Computational Statistics
  \& Data Analysis 121, 1--19.

\bibitem[{{Chao} et~al.(2015){Chao}, {H{\"a}rdle}, and {Yuan}}]{Chao:2015b}
{Chao}, S.-k., {H{\"a}rdle}, W.~k., {Yuan}, M., 2015. {Factorisable Sparse Tail
  Event Curves}. Arxiv E-prints .

\bibitem[{Cho et~al.(2014)Cho, Van~Merri{\"e}nboer, Bahdanau, and
  Bengio}]{cho2014properties}
Cho, K., Van~Merri{\"e}nboer, B., Bahdanau, D., Bengio, Y., 2014. On the
  properties of neural machine translation: Encoder-decoder approaches. Arxiv
  Preprint Arxiv:1409.1259 .

\bibitem[{Clements et~al.(2016)Clements, Hurn, and
  Li}]{clements2016forecasting}
Clements, A.~E., Hurn, A., Li, Z., 2016. Forecasting day-ahead electricity load
  using a multiple equation time series approach. European Journal Of
  Operational Research 251, 522--530.

\bibitem[{Cottet and Smith(2003)}]{cottet2003bayesian}
Cottet, R., Smith, M., 2003. Bayesian modeling and forecasting of intraday
  electricity load. Journal Of The American Statistical Association 98,
  839--849.

\bibitem[{Darbellay and Slama(2000)}]{darbellay2000forecasting}
Darbellay, G.~A., Slama, M., 2000. Forecasting the short-term demand for
  electricity: Do neural networks stand a better chance? International Journal
  Of Forecasting 16, 71--83.

\bibitem[{Dempster et~al.(1977)Dempster, Laird, and
  Rubin}]{dempster1977maximum}
Dempster, A.~P., Laird, N.~M., Rubin, D.~B., 1977. Maximum likelihood from
  incomplete data via the em algorithm. Journal Of The Royal Statistical
  Society: Series B (methodological) 39, 1--22.

\bibitem[{Dickey and Fuller(1979)}]{dickey1979distribution}
Dickey, D.~A., Fuller, W.~A., 1979. Distribution of the estimators for
  autoregressive time series with a unit root. Journal Of The American
  Statistical Association 74, 427--431.

\bibitem[{Dickey and Fuller(1981)}]{dickey1981likelihood}
Dickey, D.~A., Fuller, W.~A., 1981. Likelihood ratio statistics for
  autoregressive time series with a unit root. Econometrica: Journal Of The
  Econometric Society pp. 1057--1072.

\bibitem[{Diebold and Mariano(2002)}]{diebold2002comparing}
Diebold, F.~X., Mariano, R.~S., 2002. Comparing predictive accuracy. Journal Of
  Business \& Economic Statistics 20, 134--144.

\bibitem[{Do et~al.(2016)Do, Lin, and Moln{\'a}r}]{do2016electricity}
Do, L. P.~C., Lin, K.-h., Moln{\'a}r, P., 2016. Electricity consumption
  modelling: A case of germany. Economic Modelling 55, 92--101.

\bibitem[{Eirola and Lendasse(2013)}]{10.1007/978-3-642-41398-8_15}
Eirola, E., Lendasse, A., 2013. Gaussian mixture models for time series
  modelling, forecasting, and interpolation. In: Tucker, A., H{\"o}ppner, F.,
  Siebes, A., Swift, S. (eds.), {\em Advances In Intelligent Data Analysis
  Xii\/}, Springer Berlin Heidelberg, Berlin, Heidelberg, pp. 162--173.

\bibitem[{Elliott and M{\"u}ller(2007)}]{elliott2007confidence}
Elliott, G., M{\"u}ller, U.~K., 2007. Confidence sets for the date of a single
  break in linear time series regressions. Journal Of Econometrics 141,
  1196--1218.

\bibitem[{Elliott and M{\"u}ller(2014)}]{elliott2014pre}
Elliott, G., M{\"u}ller, U.~K., 2014. Pre and post break parameter inference.
  Journal Of Econometrics 180, 141--157.

\bibitem[{Guan et~al.(2012)Guan, Luh, Michel, Wang, and
  Friedland}]{guan2012very}
Guan, C., Luh, P.~B., Michel, L.~D., Wang, Y., Friedland, P.~B., 2012. Very
  short-term load forecasting: Wavelet neural networks with data pre-filtering.
  Ieee Transactions On Power Systems 28, 30--41.

\bibitem[{H{\"a}rdle et~al.(2012)H{\"a}rdle, M{\"u}ller, Sperlich, and
  Werwatz}]{Haerdle:2012:SPM}
H{\"a}rdle, W.~K., M{\"u}ller, M., Sperlich, S., Werwatz, A., 2012.
  Nonparametric And Semiparametric Models. Springer Science \& Business Media.

\bibitem[{Hochreiter and Schmidhuber(1997)}]{hochreiter1997long}
Hochreiter, S., Schmidhuber, J., 1997. Long short-term memory. Neural
  Computation 9, 1735--1780.

\bibitem[{Kwiatkowski et~al.(1992)Kwiatkowski, Phillips, Schmidt, and
  Shin}]{kwiatkowski1992testing}
Kwiatkowski, D., Phillips, P.~C., Schmidt, P., Shin, Y., 1992. Testing the null
  hypothesis of stationarity against the alternative of a unit root: How sure
  are we that economic time series have a unit root? Journal Of Econometrics
  54, 159--178.

\bibitem[{L{\'o}pez~Cabrera and Schulz(2017)}]{Schulz:2014}
L{\'o}pez~Cabrera, B., Schulz, F., 2017. Forecasting generalized quantiles of
  electricity demand: A functional data approach. Journal Of The American
  Statistical Association 112, 127--136.

\bibitem[{Mclachlan and Peel(2004)}]{mclachlan2004finite}
Mclachlan, G., Peel, D., 2004. Finite Mixture Models. John Wiley \& Sons.

\bibitem[{O{\u{g}}cu et~al.(2012)O{\u{g}}cu, Demirel, and
  Zaim}]{ougcu2012forecasting}
O{\u{g}}cu, G., Demirel, O.~F., Zaim, S., 2012. Forecasting electricity
  consumption with neural networks and support vector regression.
  Procedia-social And Behavioral Sciences 58, 1576--1585.

\bibitem[{Pardo et~al.(2002)Pardo, Meneu, and Valor}]{Pardo_2002}
Pardo, A., Meneu, V., Valor, E., 2002. Temperature and seasonality influences
  on spanish electricity load. Energy Economics 24, 55--70.

\bibitem[{Pesaran and Timmermann(2005)}]{pesaran2005small}
Pesaran, M.~H., Timmermann, A., 2005. Small sample properties of forecasts from
  autoregressive models under structural breaks. Journal Of Econometrics 129,
  183--217.

\bibitem[{Ramsay and Silverman(2007)}]{Ramsay:2007}
Ramsay, J.~O., Silverman, B.~W., 2007. Applied Functional Data Analysis:
  Methods And Case Studies. Springer.

\bibitem[{Rumelhart et~al.(1985)Rumelhart, Hinton, and
  Williams}]{rumelhart1985learning}
Rumelhart, D.~E., Hinton, G.~E., Williams, R.~J., 1985. Learning internal
  representations by error propagation. Tech. rep.

\bibitem[{Rumelhart et~al.(1988)Rumelhart, Hinton, Williams, and
  Others}]{rumelhart1988learning}
Rumelhart, D.~E., Hinton, G.~E., Williams, R.~J., Others, 1988. Learning
  representations by back-propagating errors. Cognitive Modeling 5, 1.

\bibitem[{Ruud(1991)}]{ruud1991extensions}
Ruud, P.~A., 1991. Extensions of estimation methods using the em algorithm.
  Journal Of Econometrics 49, 305--341.

\bibitem[{Schnabel and Eilers(2009)}]{Schnabel:2009}
Schnabel, S.~K., Eilers, P.~H., 2009. Optimal expectile smoothing.
  Computational Statistics \& Data Analysis 53, 4168--4177.

\bibitem[{Sobotka et~al.(2013)Sobotka, Kauermann, Waltrup, and
  Kneib}]{Sobotka:2013}
Sobotka, F., Kauermann, G., Waltrup, L.~S., Kneib, T., 2013. On confidence
  intervals for semiparametric expectile regression. Statistics And Computing
  23, 135--148.

\bibitem[{Stock and Watson(1996)}]{stock1996evidence}
Stock, J.~H., Watson, M.~W., 1996. Evidence on structural instability in
  macroeconomic time series relations. Journal Of Business \& Economic
  Statistics 14, 11--30.

\bibitem[{Taylor(2003)}]{taylor2003short}
Taylor, J.~W., 2003. Short-term electricity demand forecasting using double
  seasonal exponential smoothing. Journal Of The Operational Research Society
  54, 799--805.

\bibitem[{Tran et~al.(2018)Tran, Burdejov{\'a}, Ospienko, and
  H{\"a}rdle}]{Tran:2018}
Tran, N.~M., Burdejov{\'a}, P., Ospienko, M., H{\"a}rdle, W.~K., 2018.
  Principal component analysis in an asymmetric norm. Journal Of Multivariate
  Analysis .

\bibitem[{Weber et~al.(2017)Weber, Puddu, and Pacheco}]{Weber_2017}
Weber, S., Puddu, S., Pacheco, D., 2017. Move it! how an electric contest
  motivates households to shift their load profile. Energy Economics 68,
  255--270.

\end{thebibliography}

\end{document}